\documentclass[%
 reprint,
superscriptaddress,
 amsmath,amssymb,
 aps,
]{revtex4-2}
\usepackage{graphicx}
\usepackage{dcolumn}
\usepackage{nicematrix,tikz}
\usepackage{hyperref}
\usepackage{soul}

\usepackage{graphicx}
\usepackage{subfig}
\usepackage{booktabs}
\usepackage[export]{adjustbox}
\usepackage{ragged2e}
\usepackage{bm}
\usepackage{xcolor}

\let\OLDthebibliography\thebibliography
\renewcommand\thebibliography[1]{
  \OLDthebibliography{#1}
  \setlength{\parskip}{0pt}
  \setlength{\itemsep}{2pt}
}
\begin{document}

\preprint{APS/123-QED}

\title{Simulating fluid flows with quantum computing}

\author{Sachin S. Bharadwaj}
\email{sachin.bharadwaj@nyu.edu}
 \affiliation{Department of Mechanical and Aerospace Engineering, New York University, New York 11201 USA}
 
\author{Katepalli R. Sreenivasan}
\email{katepalli.sreenivasan@nyu.edu}
 \affiliation{Department of Mechanical and Aerospace Engineering, New York University, New York 11201 USA}
\affiliation{Courant Institute of Mathematical Sciences, New York University, New York, NY 10012}
 \affiliation{Department of Physics, New York University, New York, NY 10012}

\date{\today}

\begin{abstract}
\noindent The applications and impact of high fidelity simulation of fluid flows are far-reaching. They include settling some long-standing and fundamental questions in turbulence. However, the computational resources required for such efforts are extensive. Here, we explore the possibility of employing the recent computing paradigm of quantum computing to simulate fluid flows. The lure of this new paradigm is the potentially exponential advantage in memory and speed, in comparison with classical computing. This field has recently witnessed a considerable uptick in excitement and contributions. In this work, we give a succinct discussion of the progress made so far, with focus on fluid flows, accompanied by an enumeration of challenges that require sustained efforts for progress. Quantum computing of fluid flows has a promising future, but the inherently nonlinear nature of flows requires serious efforts on resolving various bottlenecks, and on synthesising progress on theoretical, numerical and experimental fronts. We present certain critical details that have not yet attracted adequate attention.\footnote{To appear in \textit{Sādhanā}, Ind. Acad. Sci. (Springer)}

\end{abstract}
\maketitle
\section{Introduction}
\label{sec:Intorduction}
Fluid flows have been intriguing for many centuries because not only of their ubiquitous presence but also of the enormous intuition they help build. Experiment and theory have been central to uncovering the phenomena most of the time, but the advent of computational technology has contributed significantly to the progress in recent years. In fact, computation has become an irreplaceable tool now, especially for consolidating the fundamental physics of complex features such as turbulence \cite{schumacher2014small,iyer2021area,buaria2023forecasting}. In the present work, we focus on an emerging computational paradigm called quantum computing, and discuss its significance for fluid dynamics research. We keep as background reference the phenomenon of turbulence, especially given its computational demands. Turbulence is supposed to be described by the Navier-Stokes equations
\begin{align}
        \frac{\partial \textbf{u}}{\partial t} + \mathbf{u}\cdot\nabla \mathbf{u} = \frac{1}{Re} \nabla ^{2} \textbf{u} -  \nabla \textbf{p},  \label{eq:governing1}
\end{align}
\begin{equation}
    \nabla \cdot \textbf{u} = 0,
    \label{eq:incompressibility}
\end{equation}
where $\mathbf{u}$ is the velocity field, 
$\mathbf{p}$ is the pressure, $Re = {\bar{U}}L/\nu$ is the Reynolds number, ${\bar{U}}$ being the characteristic velocity, $\nu$ the kinematic viscosity and $L$ the characteristic length. The magnitude of $Re^{-1}$ quantifies the nonlinearity in the flow and the intensity of turbulence. This nonlinear problem is still unresolved in its generality for any boundary conditions. 

Let us consider the three-dimensional, direct numerical simulation of a homogeneous, isotropic turbulence problem \cite{schumacher2014small}. The typical numerical setup is a periodic box of side $L$. The equations are now solved using a pseudo-spectral method by discretizing the box into $N$ grid points in each direction, with a total number of $N^{3}$. It is important to choose an $N$ such that even the smallest length scales in the flow are well resolved. Since this smallest scale is orders of magnitude smaller than $L$ for high $Re$, the number $N$ required to represent all scales is quite large \cite{yeung2023turbulence}. Thus, simulations of high Reynolds numbers turbulence push the bounds of classical digital computing on even the most powerful computers. Thus, a different computing paradigm is required. Here, we consider the potential of quantum computing for the purpose.

Quantum computing performs mathematical operations following the rules of quantum mechanics, which endow such computers with potentially radical advantages of speed and memory (the ``quantum advantage"). How to apply quantum algorithms to solve classical problems more efficiently than classical technology? A few years ago in ref. \cite{bharadwaj2020quantum}, we made an initial proposition of potential directions, in view of initiating such a field of study. Most of these methods have now been explored to varying extents. By assimilating the lessons learnt from the progress made since then, we attempt here to provide an updated view of the (sub)field of Quantum Computation of Fluid Dynamics (QCFD). We give a brief overview of contributions made in this field so far, and outline their advantages and shortcomings. Our intent is not to make an in-depth comparison of these methods but highlight specific challenges that require focused efforts to approach the quantum advantage. We also call attention to details and caveats that generally go unnoticed but are critical to realizing actual flow simulations on a real quantum computer (with some or most of the promised quantum advantage). It is our hope that this work will serve as an up-to-date starting point for entrants to the field. 

The rest of the manuscript is organized as follows. First, we give in Section 2 a quick introduction to concepts of quantum computing and their advantages. In Section \ref{sec:challenges thinking quantumly}, we discuss the challenges posed, such as the hardships of handling nonlinearity, data input and output, noise and decoherence, all of which tend to diminish the quantum advantage. We then outline methods of addressing these issues. In Section \ref{sec:Quantum algorithms} we discuss some important quantum algorithms that have been developed to solve flow problems. We briefly comment on the veracity of quantum advantage in Section \ref{sec: veracity of advantage}, and conclude with a summary and discussion in Section \ref{sec: discussion}.


\begin{figure*}[htpb!]
    \centering
    \includegraphics[trim={0.2cm 0.1cm 0.2cm 0.2cm},clip=true,scale=0.3]{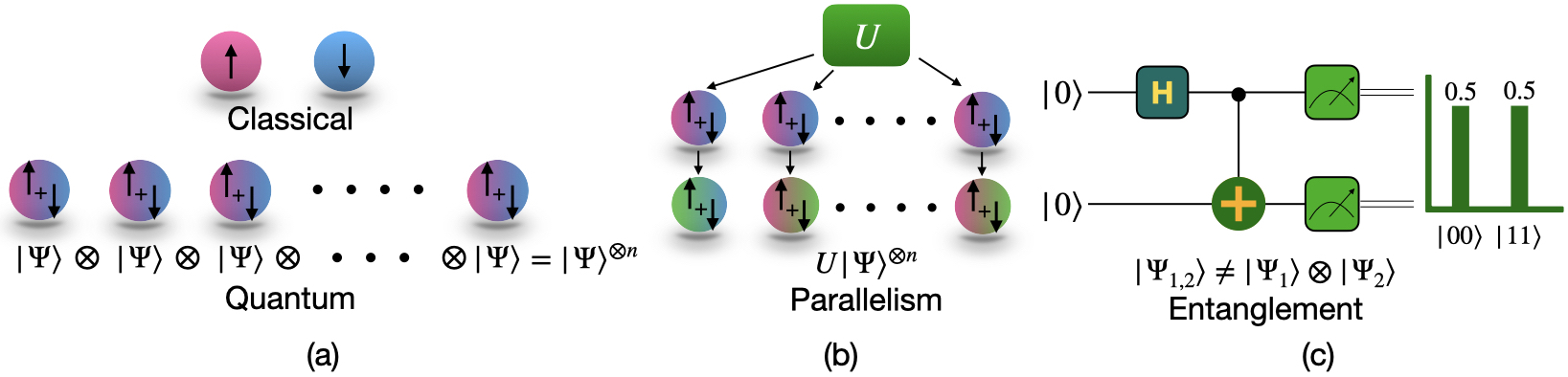}
    \caption{\justifying Schematic of the three distinguishing features of a quantum computer: {(a) shows a system of n-qubits whose collective, tensorial product space can encode exponentially larger data than classical bits of information; (b) depicts the innate parallelism of quantum operators that act on exponentially large vector spaces; (c) shows a quantum circuit that creates an entangled pair of qubits, whose basis states have a probability distribution shown by the histogram.} These concepts are explained in Section 2 of the text.}
    \label{fig:qubits_parallel_entanglement}
\end{figure*}

\section{What's different about Quantum Computing?}
\label{sec:Whats different}
Quantum computing is associated with storing and processing of information according to quantum mechanical rules. This shift in focus brings with it lucrative computational advantages as well as unfavorable bottlenecks that need to be addressed. Three core concepts, shown schematically in figure \ref{fig:qubits_parallel_entanglement}, give a quantum computer its edge and makes it fundamentally distinct from the classical ones. A comprehensive introduction may be found in \cite{nielsen2002quantum}.

\subsection{QUBITS}
\label{subsec:qubits}
The first concept is that of the memory or physical encoding of data. Classical computers store information in terms of binary bits, which are either 0 or 1 at any given instant of time. Whereas quantum computers store information using \textit{qubits}, which represent the physical state of a quantum object, denoted by the wavefunction $\vert\Psi\rangle \in \mathbb{C}^{2}$. Consider the spin of an electron. By the rules of quantum superposition, it is either spin-up ($|\uparrow\rangle = \vert 1\rangle$), spin-down ($|\downarrow\rangle = \vert 0\rangle$) or a vector superposition of ($v_{0}|\uparrow\rangle+v_{1}|\downarrow\rangle$ = $v_{0}|0\rangle+v_{1}|1\rangle$), where $v_{0},v_{1} \in \mathbb{C}$.  Physically, $v_{0}^{2}\textrm{~and~}v_{1}^{2}$ indicate the probability of finding the spin state in its corresponding basis state upon measurement. This immediately necessitates a normalized state with $\vert\vert\vert\Psi\rangle\vert\vert=\sum_{i}v^{2}_{i}=1$. The information we intend to store (in our case this could be the discretized velocity field $\mathbf{v_{i}}$) is encoded either in terms of these complex coefficients $v_{0},v_{1} \dots$ (called \textit{amplitude encoding}), or as binary values corresponding to the binary bases themselves, such as $|0\rangle,|1\rangle,|01\rangle,|11\rangle \dots$ (called \textit{basis or state encoding}). Although the former encoding seems to be a popular choice given the access to a continuous range of complex numbers, both methods find their merited places, based on the algorithm design. For most of the remaining discussion, we will assume amplitude encoding.
 
A system of $n$ qubits would therefore produce a tensorial product state, that spans a $N=2^{n}$ dimensional vector space (Hilbert space, $\mathbb{C}^{2^{n}}$), written as $\vert\Psi\rangle^{\otimes n} = \sum_{i=0}^{2^{n}-1}v_{i}\vert i\rangle$, allowing one to encode an exponentially large data set. Now consider, for instance, storing the velocity field generated by the largest, classical direct numerical simulation of a homogeneous isotropic turbulence problem \cite{yeung2023turbulence} with $N = 32768^{3}$ grid points. This would need only about $n=\log_{2}(N)=45$ qubits, while current and near-term quantum computers have in them a few hundred to a thousand qubits. Simulating a flow problem with a quantum algorithm needs far larger number of qubits than those required to just store the velocity field. These ancillary qubits are necessary to perform certain algorithm-specific operations. It is thus preferable to design algorithms that require the total number of qubits (\textit{qubit complexity}) to be logarithmic in the problem size in order to attain the exponential advantage in memory. The physical realization of these qubits assumes multiple options. Some popular ones are based on one of the following technologies: superconductors, ultra-cold atoms, ion traps, neutral atoms, semiconductor or photonics based quantum processors; each of them has its own merits and drawbacks. It is important to remember such machine capabilities while designing quantum algorithms specific to a certain hardware \cite{tennie2024quantum}.

\subsection{QUANTUM PARALLELISM}
\label{subsec:parallelism}
The second concept of quantum parallelism deals with how the stored information is processed via a quantum algorithm. A quantum algorithm is composed of a quantum circuit that acts on the qubits to perform a specific numerical operation. The quantum circuit itself is a collection of several quantum gates that can be mathematically described as unitary matrix operators $U$ (where $U^{\dag}U=UU^{\dag}=1$), acting on the qubits. Physically, these operators could be electric or magnetic fields that affect the state of a qubit in a well-defined and controllable manner. The quantum circuit is a specific organization of gate operators, deciding the time evolution trajectory of the qubit state. Such an evolution can be captured by the well-known Schr\"odinger wave-equation
\begin{equation}
    i\hbar\frac{\partial\vert\Psi\rangle}{\partial t} = H\vert\Psi\rangle,
\end{equation}
where $H$ is a Hermitian, Hamiltonian operator that describes the interaction of the qubit state with the externally applied fields. This implies that the evolving quantum state at some later time $t$ is given by the unitary transformation
\begin{equation}
    \vert\Psi(t)\rangle = e^{\frac{-iHt}{\hbar}}\vert\Psi(0)\rangle = U\vert\Psi(0)\rangle = \sum_{j=0}^{2^{n}-1}U(v_{j}\vert j\rangle).
\end{equation}
Note here that the operator $U$ acts on all the vector elements $v_{j}$ at the same time in parallel. This indicates the inherently parallelized nature of quantum gates and circuits, which can simultaneously operate on an exponentially large vector space. Quantum algorithms that efficiently utilize this property (``time complexity") exhibit an exponential speed-up. Some popular instances of this property, relevant to numerical flow simulations, could be the Quantum Fourier Transform (QFT) {\cite{nielsen2002quantum}} or the HHL algorithm {\cite{harrow2009quantum}} (named after its creators, Harrow, Hassidim and Lloyd), where the former performs a discrete Fourier transform and the latter solves a system of linear equations (``matrix inversion problem"). For a problem of size $N$, both these algorithms perform their respective tasks in $O(\log(N))$ operations, whereas their classical counterparts take at least $O(N)$ operations, thus enabling an exponential speed up in the asymptotic complexity scaling. 

\subsection{ENTANGLEMENT}
\label{subsec:entanglement}
The third concept native to quantum computers is quantum entanglement. Consider two qubits set to $\vert0\rangle$ to begin with, as shown in figure \ref{fig:qubits_parallel_entanglement}(c), which is a simple example of a quantum circuit. The horizontal lines represent the time evolution of these qubits. These qubits are then operated on by a series of three quantum gates that cause the following transformations. The number of irreducible layers of quantum gates determines the \textit{circuit depth}, while the \textit{circuit width} is simply the number of qubits. The circuit has the following characteristics:
\begin{itemize}
    \item First, a Hadamard gate \textbf{H} (which turns a state of $\vert 0\rangle$ or $\vert 1\rangle$ into an equal superposition of $\vert 0\rangle$ and $\vert 1\rangle$)
 is applied on the first qubit which gives
    \begin{equation}
        \vert0\rangle\otimes\vert0\rangle \xrightarrow{H\otimes I} \frac{\vert0\rangle + \vert1\rangle}{\sqrt{2}}\otimes\vert0\rangle = \frac{1}{\sqrt{2}}(\vert00\rangle + \vert10\rangle).
    \end{equation}
    \item Next, a controlled-NOT (CNOT) gate is applied on the second qubit. This flips the target qubit conditioned on whether or not the control qubit (qubit 1) is set to $\vert1\rangle$, thus giving,
    \begin{equation}
         \frac{1}{\sqrt{2}}(\vert00\rangle + \vert10\rangle)\xrightarrow{CNOT}\frac{1}{\sqrt{2}}(\vert00\rangle + \vert11\rangle).
         \label{eq:Bell state}
    \end{equation}
    \item Finally, we apply the measurement operators on both qubits. Repeating the circuit and conducting multiple measurements gives a probability distribution (histogram) for the basis states. From the above equation, it is clear that of the four possible basis states, $\vert00\rangle$ and $\vert11\rangle$ have equal and finite probabilities of 0.5 each, as shown to the extreme right of figure 1.
\end{itemize}
The circuit described above creates a state called a Bell state, which is a quantum entangled state. The state in eq.~(\ref{eq:Bell state}) is entangled in the sense that it can no longer be expressed as a tensor product of two separate qubit states \big($\vert\Psi_{1,2}\rangle\neq\vert\Psi_{1}\rangle\otimes\vert\Psi_{2}\rangle$\big). This property causes a correlation between the different vector subspaces that not only connects together all the information and is manipulable because of it. At the same time, it becomes hard to work on an independent subspace without affecting the others. Thus, if one can devise ways to use entanglement for exploring nonlinear and nonlocal transformations and observables such as velocity correlation functions, quantum setting becomes natural for flow computations \cite{gourianov2022quantum}.

The above three concepts form the core of quantum algorithm design, whose advantages we intend to harness for realizing efficient quantum simulations of fluid flows. The gains, however, are accompanied by certain bottlenecks. We discuss a few key ones in the following section, and also address how to maintain quantum advantage while successfully solving practical flow problems on current or near-term quantum hardware.

\section{Challenges that demand ``thinking quantumly"}
\label{sec:challenges thinking quantumly}
The field of Computational Fluid Dynamics (CFD) \cite{anderson1995} has evolved greatly since the 1960's. It now hosts a plethora of high-performance algorithms to solve flow problems numerically. 
The first natural step of quantum computing is to replace a computationally expensive operation in these algorithms (perhaps the entire algorithm itself) by a quantum alternative that performs the same task more efficiently. It is an open question as to whether this is the most successful approach to take. Nevertheless, there seems to have been some considerable progress in this direction in the recent past, which include algorithms that offer up to an exponential advantage compared to classical choices, under certain important caveats, some of which we outline briefly in Section \ref{sec:Quantum algorithms}. 
But first we address the challenges that quantum computers face. In essence, these challenges motivate the need to \textit{think quantumly}, in order to design and develop new algorithms that can outperform classical ones.  

The major bottlenecks arise from the linear nature of quantum mechanics, which poses a challenge for handling nonlinear problems. We highlight them by means of a typical hybrid quantum-classical algorithm workflow by assuming access to an \textit{ideal} quantum computer with no hardware inaccuracies. The limitations imposed by the currently available quantum hardware are considered in Section 3.4. 


\subsection{HYBRID QUANTUM-CLASSICAL ALGORITHMS} 
\label{subsec:hybrid algorithms}
\begin{figure}[htpb!]
    \centering
    \includegraphics[trim={0.2cm 0.2cm 0.1cm 0.2cm},clip=true,scale=0.34]{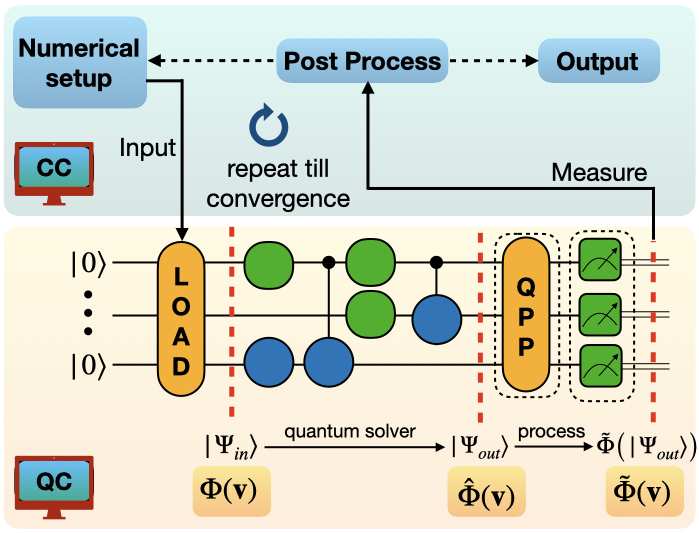}
    \caption{\justifying Schematic of the workflow of a generic hybrid quantum-classical algorithm. {The classical and quantum computers are denoted by CC and QC, respectively. 
    The classical device first sets up the problem numerically and the essential information is encoded into the quantum device, denoted as $\Phi(\textbf{v})$. The quantum device performs a specific information processing task, producing a state $\hat{\Phi}(\textbf{v})$, which is either read-out (measured) directly or post-processed quantumly. We denote this final state by $\tilde{\Phi}(\textbf{v})$, which is then communicated back to the classical device for further processing.} }
    \label{fig:hybrid_workflow}
\end{figure}

The typical workflow of a hybrid algorithm is shown in figure \ref{fig:hybrid_workflow}. First, the numerical setup is constructed classically (the part within the block CC). 
This could include developing a matrix equation from a finite difference method, or a parameterised state for an optimization based problem. Once these objects are prepared, the input data (e.g., the initial condition of the flow field) is encoded using a quantum state preparation algorithm, and a quantum state $\vert\Psi_{in}\rangle$ is created. 
The quantum computer then executes the core algorithm, performs a specific operation on this input state, and produces the output state $\vert\Psi_{out}\rangle$ (the block QC). 
At this point, we may choose to either perform a full measurement of the state (quantum state tomography) by outputting the entire velocity field, or performing quantum post processing to compute a function 
of the final state, which is measured subsequently. This could even be a single real-valued number computed from the output (as in the days of old analogue experiments). In any case, the measured information is read into the classical computer. Based on the overall algorithm, one might need to use this result to iterate on the same steps until convergence is attained. It is the complexity of this overall process that needs to scale better than that of a purely classical algorithm.

\subsection{DATA ENCODING AND MEASUREMENTS}
\label{subsec:data encoding and measurments}
Quantum state preparation (data encoding) and state measurements are necessary operations for the communication between classical and quantum devices. However, in the most general case, the computational complexity of these input-output operations tends to diminish the quantum advantage of the overall algorithm and is the first critical bottleneck. For specificity, we illustrate the problem for a simple input representation of discretized velocity field encoded in terms of amplitudes as $\Phi(\mathbf{v})\equiv\sum_{i}v_{i}|i\rangle$.

\begin{figure}[htpb!]
    \centering
    \includegraphics[trim={0.2cm 0.2cm 0.1cm 0.2cm},clip=true,scale=0.35]{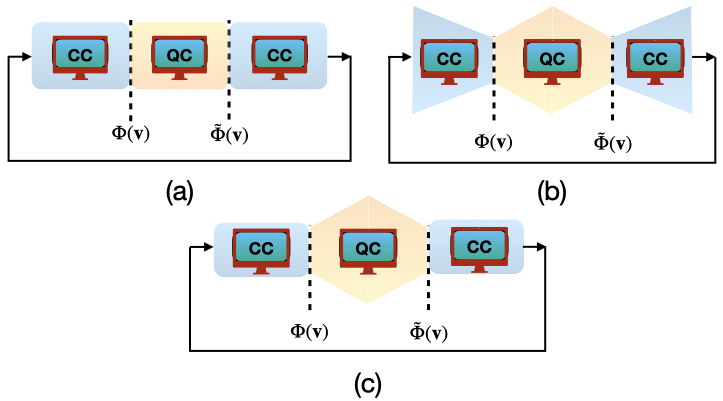}
    \caption{\justifying Schematics of different data flow volumes between the {classical (CC) and quantum computers (QC). (a) depicts the case where the classical and the quantum computers process and communicate equal data volumes. In (b), although both devices process equal data volumes, the data undergo compression during communication. (c) shows the case where the data being communicated and processed by the classical device are always smaller than the data being processed quantumly.}}
    \label{fig:input_output_flow}
\end{figure}

(a) \textit{Storage mismatch --} 
    Consider the 1121 qubit quantum device (IBM Condor \cite{choi2023ibm}). Even if one used only 100 qubits to store the velocity field, it would correspond to about $\approx 10^{30}$ values or $N^{3}=(10^{10})^{3}$ grid points for turbulence simulations. This is on the order of a million times bigger than the largest classical simulation! However, this would require about $\approx 10^{12}$ exabytes of classical storage, which is a clear impossibility at present. Therefore, even though a quantum device has an exponentially large memory, the data sizes being communicated in and out of a quantum computer, as shown in figure \ref{fig:input_output_flow}(a), constitute a limitation. This situation triggers the following question: Can we process only as much data quantumly as is allowed by classical storage capacities? 
    
(b) \textit{Complexity bottlenecks --} Encoding a general data vector $\Phi(\mathbf{v})$ into qubits calls for quantum circuits whose depth generally scales linearly with the data size as $\mathcal{O}(N)$. For a quantum encoded data of size $N$, measuring the entire quantum state is an $\mathcal{O}(N)$ operation as well. This is so since one would have to re-prepare the quantum state at least $\mathcal{O}(N)$ times, since the act of measuring even a single data element causes the loss of the remaining information. However, specific data representations can be encoded rather efficiently as we shall shortly point out. 

The above discussion clarifies that the forms of $\Phi(\mathbf{v})$ and $\tilde{\Phi}(\mathbf{v})$ need to be chosen differently to avail oneself of the quantum advantage, and they should carefully consider efficient state preparation and measurement. To ensure that quantum advantage in time complexity is preserved, one needs to either choose specific forms of $\Phi(\mathbf{v})$ and $\tilde{\Phi}(\mathbf{v})$, for which there exist efficient quantum circuits, or
control the amount of data at the entry and exit of the quantum device. This latter hints at possible data compression required at both input and output stages, such that the circuit complexity is kept lower than $\mathcal{O}(N)$. The data could, however, be decompressed anywhere within the classical or quantum device, except during communication as shown in figure \ref{fig:input_output_flow}(b).

In addition to controlling data flow at input and output, it is also desirable to design algorithms such that the classical device always handles a lower data load than the quantum device at all times, as shown in figure \ref{fig:input_output_flow}(c). This will ensure that the quantum advantage of memory is always preserved.


(c) \textit{Measurement Errors --} We recall that the amplitudes of a given quantum state are essentially elements of a probability distribution. Its measurement implies reconstructing this function by querying and sampling it multiple times. Besides the fact that measurement operations are themselves imperfect, insufficient sampling contributes to additional statistical errors, which decays according to the Law of Large Numbers as $1/\sqrt{N_{s}}$, where $N_{s}$ is the number of samples or measurements \cite{bharadwaj2024compact}. Suitable measures need to be taken such that this error is kept under control.

\subsection{NONLINEARITY}
\label{subsec:nonlinearity}
The linear nature of quantum mechanics presents another major challenge for solving practical flow problems, which are almost always nonlinear. To understand this challenge, 
note that any numerical scheme for solving nonlinear flow equations requires the computation of the nonlinear quantity $u^{2}$---that is, the multiplication of the velocity with itself multiple times. However, two consequences of quantum mechanical principles are: 
\begin{itemize}
    \item \textit{Multiplication is hard} -- Quantum circuits can only perform unitary and linear transformations on the target variable and therefore computing nonlinear transformations of the type $u\xrightarrow{U=?} u^{2}$ using a quantum device is non-trivial.
    \item \textit{Copying is a No-Go} -- Any numerical scheme would need to make at least a single copy of the variable in its original form $u$ at every time step. However, the No-Cloning theorem \cite{wootters1982single} prohibits us from copying any information within a quantum algorithm.
\end{itemize}
These principles emphasize the need for coding and processing the velocity data in ways that are more natural for a quantum processor. These blockades are mostly algorithmic. We will now outline challenges that need to be addressed while implementing algorithms on real quantum hardware.




\subsection{CHALLENGES WITH REAL QUANTUM DEVICES}
\label{subsec:challeneges with real devices}
While the exponential advantage in memory due to qubits seems propitious, we should note the quantum information storage (at least on currently available devices) is more short-lived \cite{ganjam2024surpassing} than even the classical Random Access Memory (RAM). Furthermore, qubits are highly sensitive to interactions with the classical environment, or even minor (unwanted) external fluctuations, which express themselves as \textit{decoherence} and \textit{noise} in quantum information stored in qubits, and degrade the overall accuracy of a quantum algorithm \cite{suter2016colloquium,naghiloo2019introduction}. 

\begin{figure}[htpb!]
    \centering
    \includegraphics[trim={0.1cm 0cm 0.1cm 0.2cm},clip=true,scale=0.26]{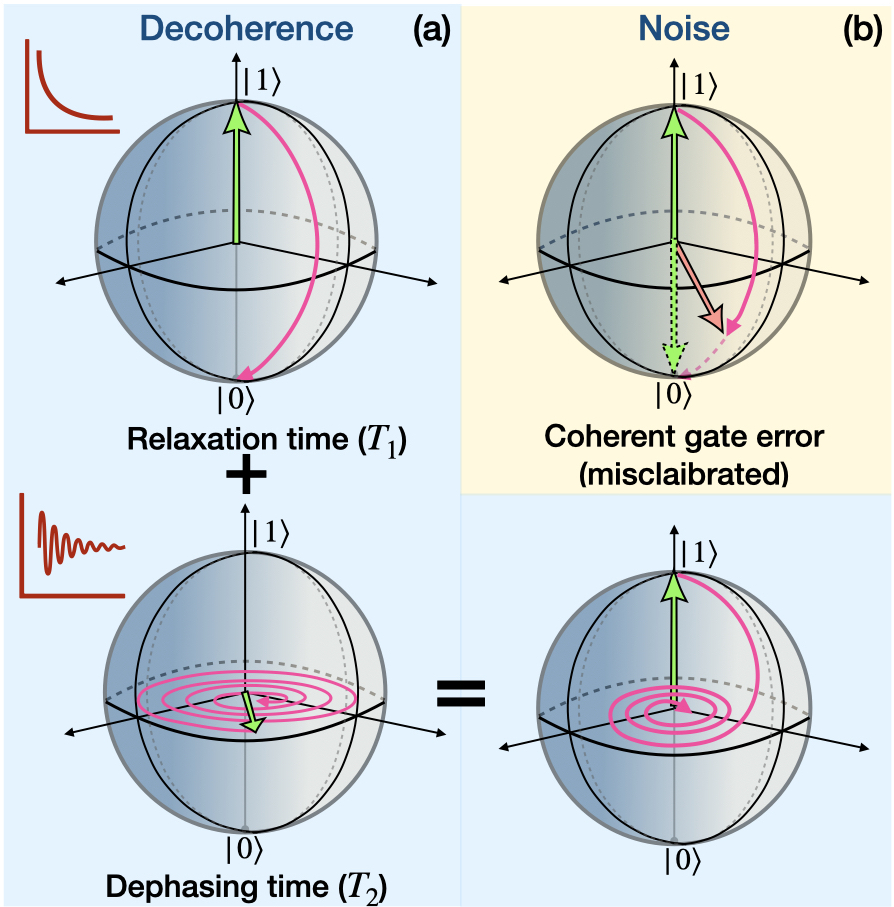}
    \caption{\justifying Schematic visualizing the sources of error on real quantum hardware on a Bloch sphere, from (a) Decoherence relaxation ($T_{1}$) and dephasing ($T_{2}$) times, based decoherence of qubits and (b) Gate error based noise in quantum circuits.}
    \label{fig:T1_T2_Bloch}
\end{figure}

(a) \textit{Decoherence --}
  A general quantum state, visualized as a vector on a Bloch sphere, as shown in figure \ref{fig:T1_T2_Bloch}, is given by
 \begin{equation}
    |\Psi\rangle = e^{i\beta_{1}}\Big[\cos\beta_{2}\|0\rangle + e^{i\beta_{3}}\sin\beta_{2}|1\rangle \Big]\label{eq:bloch state},
\end{equation}
where $\beta_{1}$ is the global phase,  $\beta_{2}$ is the polar angle and $\beta_{3}$ the relative phase. Suppose a qubit is prepared in an excited state $\vert1\rangle$ as shown in the top-left panel of figure \ref{fig:T1_T2_Bloch}(a). Even if one isolates a qubit from external interactions and noise, given some time $\tilde{t}$, the qubit state can relax into the ground state $\vert0\rangle$ or some mixed state, thus losing the original information stored. This decoherence mechanism is called \textit{qubit relaxation}. The population of the excited state that decays exponentially into a ground/mixed state follows the relation $p(\tilde{t})=p(0)e^{-\tilde{t}/T_{1}}$. The second mechanism of data corruption, called \textit{dephasing}, is a consequence of external noise by which a qubit begins to stochastically rotate in the Bloch sphere, distorting the phase information stored in eq.~(\ref{eq:bloch state}). This is shown in the bottom-left panel of figure \ref{fig:T1_T2_Bloch}(a). The dephasing is also an exponentially decaying function with a corresponding time constant $T_{2}$. In most cases these two mechanisms occur simultaneously, as shown in the bottom-right panel of figure \ref{fig:T1_T2_Bloch}(a). It is thus important to design short and compact quantum circuits whose total execution time is shorter than $\min\{T_{1},T_{2}\}$. Typically, these constants are of the order of $\sim 100\mu s$, while more recent experiments seem to march beyond \cite{ganjam2024surpassing}. In parallel, current experimental efforts are aimed at obtaining new physical realizations of a qubit that are more robust against these limitations. 

(b) \textit{Gate noise and error rates --}
Coherent gate errors due to miscalibration form a key contributor to quantum noise. Consider a simple \textbf{X} gate (NOT gate) acting on a state initially prepared as $\vert1\rangle$. In the ideal case, the outcome is $\textbf{X}\vert1\rangle=\vert0\rangle$. This would be represented as a full $180^{\circ}$ flip in the Bloch vector as shown in figure \ref{fig:T1_T2_Bloch}(b). However, due to a faulty gate action, sometimes the state would be flipped to some point in between, producing a small error which amounts to noise in the final solution. Based on the number of gates a quantum circuit requires, such gate errors accumulate producing erroneous solutions. This is generally characterized by the \textit{error rate}, which is the probability with which a gate would fail to produce the desired result. Most commercially  accessible quantum devices currently have error rates of about $10^{-3}$. 
This necessitates extremely compact quantum circuit designs, with a small enough total gate count. These numbers suggest that gate counts on current devices might be limited to only about $\mathcal{O}(100)$ one- and two-qubit gates \cite{bharadwaj2024compact}. 

%

\section{Quantum algorithms for flow simulations: a glimpse}
\label{sec:Quantum algorithms}
We now describe briefly the different quantum computational fluid dynamics algorithms and approaches proposed so far. Our goal is to provide a perspective of areas with potential for applications, deserving of focused efforts. We first list the proposed techniques for dealing with flow nonlinearities.

\subsection{DEALING WITH NONLINEARITY}
\label{subsec:dealing with nonlinearity}
\begin{figure}[htpb!]
    \centering
    \includegraphics[trim={0.1cm 0cm 0.1cm 0.2cm},clip=true,scale=0.22]{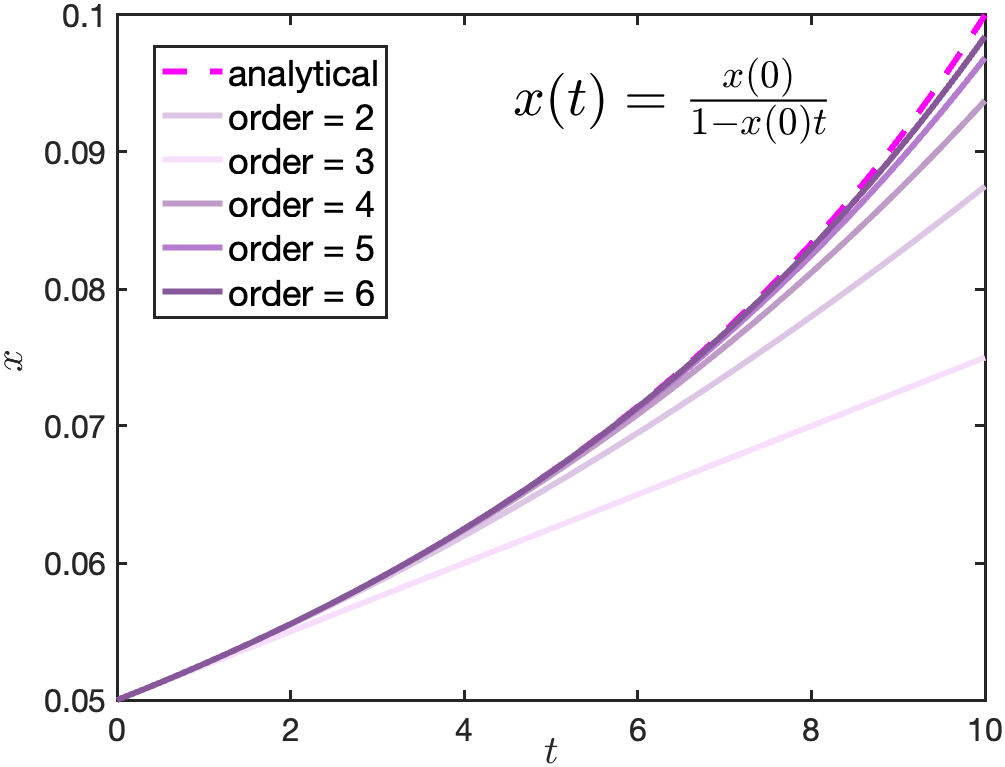}
    \caption{\justifying Solid lines of increasing thickness represent increasing orders of Carleman truncation, whereas the dotted line shows the analytical result.}
    \label{fig:carleman order}
\end{figure}

The direct approach for dealing with nonlinearity is to simply begin by constructing a quantum circuit with multiple copies of the state. {A tensorial product of states containing multiple copies of the original data automatically encodes all higher order nonlinear terms.} This would evade the blockades from the no-cloning theorem and nontrivial operations such as nonlinear transforms with quantum gates. However, as shown in ref. \cite{leyton2008quantum}, such a strategy would be inefficient and the number of qubits required would grow exponentially with the degree of nonlinearity and the dimensionality of the problem. This reaffirms the need for an alternative encoding $\Phi(\bf{v})$. Techniques proposed in this light are: 
  
    (a) \textit{Mapping a finite dimensional nonlinear system into an \textit{infinite} dimensional linear system} --  Carelman, Koopman and Homotopy methods are some of the techniques employed for this purpose \cite{liu2021efficient,engel2021linear,lin2022koopman,giannakis2022embedding,bharadwaj2023quantum,itani2024quantum}. Consider as an example a simple quadratic nonlinearity of the form $\dot{x}=x^{2}$. Given an initial condition $x(0)$, this equation has the analytical solution given by $x(t)=x(0)/(1-x(0)t)$. Let us now perform a mapping into a new set of variables $y$ as follows. Let us denote $x=y_{1}$ and therefore $\dot{y}_{1}=x^{2}$. Now let us set $x^{2}=y_{2}$ and the equation of motion for this new variable would be $\dot{y}_{2}=2x^{3}$. We can carry out this process similarly and construct a large system of linear ODEs for the new set of variables, defined for every high-order term. Such a series of equations is generally truncated to some order. This would give a matrix equation of the form
    \begin{align}
        \frac{d}{dt}\begin{pmatrix}
            y_{1}\\y_{2}\\y_{3}\\ \vdots 
        \end{pmatrix} = &\begin{pmatrix}
            0 & 1 & 0 & 0 &  \cdots & \\
            0 & 0 & 2 & 0 &  \cdots & \\
            0 & 0 & 0 & 3 & 0  & \cdots\\
            0 & 0 & 0 & 0 & 4 & \cdots  
        \end{pmatrix}\begin{pmatrix}
            y_{1}\\y_{2}\\y_{3}\\ \vdots 
        \end{pmatrix} \nonumber \\
        &\frac{dy}{dt} = My 
    \end{align}
    It can be seen from figure \ref{fig:carleman order} that increasing orders of truncation makes the numerical Carleman solution to closely follow the analytical result\cite{liu2021efficient,sanavio2024three}. It also reveals that the accuracy for lower truncation orders is maintained up to shorter windows of time evolution. Quantum algorithms that begin with continuum {or meso-scale} formulations of fluid flows typically resort to this approach, {also known as \textit{linearization or linear embedding}}.
    
    (b) \textit{Binary encoding} -- An alternative strategy used in certain Lattice Boltzmann methods {(a meso-scale approach)} \cite{itani2024quantum,bharadwaj2023hybrid}, and in Quantum Post Processing algorithms \cite{mitarai2019quantum,rattew2023non,bharadwaj2023hybrid}, is to encode the data in binary format using the basis of quantum states. For instance, the set of basis states of a three qubit system could map to the following set of values.
    \begin{align}
        \textrm{Basis state} ~~~~ & \textrm{position index} ~~~\textrm{fractional binary} \nonumber \\ 
        \vert000\rangle \rightarrow & \hspace{1cm}0 \hspace{2cm} 0 \nonumber \\
        \vert001\rangle \rightarrow & \hspace{1cm}1 \hspace{2cm} 0.125 \nonumber \\
        \vert010\rangle \rightarrow & \hspace{1cm}2 \hspace{2cm} 0.25 \nonumber \\
        \vert011\rangle \rightarrow & \hspace{1cm}3 \hspace{2cm} 0.375 \nonumber \\
        \vert100\rangle \rightarrow & \hspace{1cm}4 \hspace{2cm} 0.5 \nonumber \\
        \vert101\rangle \rightarrow & \hspace{1cm}5 \hspace{2cm} 0.625 \nonumber \\
        \vert110\rangle \rightarrow & \hspace{1cm}6 \hspace{2cm} 0.75 \nonumber \\
        \vert111\rangle \rightarrow & \hspace{1cm}7 \hspace{2cm} 0.875 \nonumber 
    \end{align}
The Lattice Boltzmann approaches use such an encoding to represent locations of particles or number of particles. The nonlinear interactions are then performed as bit-operations that involve a sequence of CNOT gates. The same basis states could also represent fractional binary numbers as shown above and, by using controlled rotation or bit-arithmetic operations \cite{mitarai2019quantum,bharadwaj2023hybrid}, nonlinear transforms of these values can be computed. In both cases, since all the data are represented using finite precision of binary basis states, the accuracy of the method critically depends on the number of qubits. 
    
\begin{figure}[htpb!]
    \centering
    \includegraphics[trim={0.1cm 0cm 0.1cm 0.2cm},clip=true,scale=0.35]{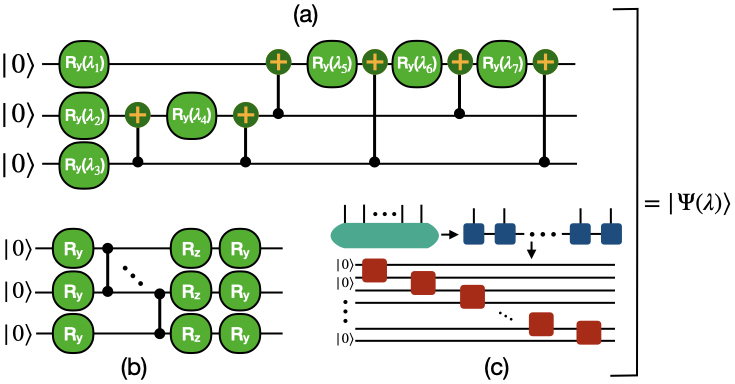}
    \caption{\justifying (a) shows the universal ansatz applied on 3 qubits. It can accurately parameterize and reproduce any real valued input. However, since the required parameters scale linearly with the input data size $N$, optimizing them can be computationally expensive. To circumvent this, one could resort to a hardware-aware ansatz, as shown in (b), offering shallow circuit depths, customized to specific qubit topologies. A further improvement of these ansatzes is made with ADAPT(ive), optimal ansatzes as discussed in ref.\cite{grimsley2019adaptive}. (c) shows an alternative, robust method to encode input data via an MPS ansatz.}
    \label{fig:ansatzes and mps}
\end{figure}

(c) \textit{Other ansatzes} -- Instead of an exact encoding of quantum amplitudes, one could also consider a parametrization of the input data as an initial approximation, which can be iteratively improved upon to match the target state. This is a technique primarily used in context of Variational Quantum Algorithms \cite{cerezo2021variational}. A given state $\vert\Psi\rangle$ can be approximated by an ansatz state $\vert\tilde{\Psi}(\bf{\lambda})\rangle$, where $\bf{\lambda}$ is a set of parameters. The number of parameters is obviously expected to be smaller than the size of the dataset itself. This input ansatz state $\vert\tilde{\Psi}(\bf{\lambda})\rangle$ is prepared by a fixed quantum ansatz circuit $\tilde{U}(\lambda)$ that is generally a sequence of rotation ($R_{y}$) and CNOT gates. Ideally, one would expect low depth circuits (to alleviate state preparation overheads), hardware awareness, and reproduction of the input state with high accuracy, making sure that the last step does not lead to over- or under-parametrization issues resulting is a plateaued behavior. Examples of such an ansatz are shown in figures \ref{fig:ansatzes and mps}(a) and (b). Such ansatzes have a heuristic flavor and are customized for each problem. 

Alternatively, there also exist certain robust strategies that offer efficient ways to encode the input ansatz such as the Matrix Product States (MPS). This ansatz, also known as Tensor Train ansatz, is a tool used widely in quantum condensed matter physics, which takes as its input a tensor product state such as $$\vert\Psi\rangle = \sum\limits_{q_{1},\dots,q_{n}}T[1]^{q_{1}}\dots T[n]^{q_{n}}\vert q_{1}\rangle\otimes\dots\otimes\vert q_{n}\rangle,$$ where $T[k]$ are tensors acting on Hilbert spaces of $q_{k}$. Such a state can be first broken down into a train of more compact and smaller matrices that are connected by a single parameter set as shown in figure \ref{fig:ansatzes and mps}(c). These can be further decomposed into a sequence of only isometric operators that can be translated into a quantum circuit of unitary operators \cite{lubasch2020variational}. The MPS states in essence offer a basis of polynomial functions and Fourier series that can be used to approximate the input state as an ansatz. On similar lines one can also use certain product feature maps and Chebyshev feature maps as in \cite{kyriienko2021solving}. 

To encode nonlinearity of the kind described above, 
if the nonlinearity is of order $d_{n}$, one could make $d_{n}$ copies of the parameterized state to encode all terms of order $d_{n}$, which is the central tenet behind the proposal in ref. \cite{lubasch2020variational}. At first glace it appears to be more resource efficient than a linearization scheme, but the success of this approach critically depends on a reliable ansatz and an associated optimization scheme devoid of barren plateaus (regions where zero gradient does not correspond to local minima of interest) \cite{holmes2022connecting,larocca2024review}.

    (d) \textit{Equivalent formulations} -- Instead of solving PDEs, requiring one to handle nonlinearity explicitly, we can turn to alternative formulations.
    
    (i) The Liouville and Fokker-Planck equations \cite{lin2022koopman,succi2024ensemble}, written respectively as
    \begin{align}
        \frac{\partial\rho(\mathbf{x},t)}{\partial t} + \frac{\partial}{\partial \mathbf{x}}[\Phi(\mathbf{x}),\rho(\mathbf{x},t)] =0,\label{eq:liouville}\\
        \frac{\partial\rho(\mathbf{x},t)}{\partial t} + \frac{\partial}{\partial \mathbf{x}}[\Phi(\mathbf{x}),\rho(\mathbf{x},t)] = D\frac{\partial^{2}}{\partial \mathbf{x}^{2}}[\rho(\mathbf{x},t)],\label{eq:fokker planck}
    \end{align}
fall into this category, called master equations. Here, $\rho(\mathbf{x},t)$ is the probability distribution of the state of a given nonlinear dynamical system that evolves in space and time and the above equations capture its time evolution. $\Phi(\mathbf{x})$ is a specific flow process with which the system evolves, $D$ is the diffusion constant and $[~.~]$ represents a quadratic variation. Notice that both these equations are liner (though they represent an underlying nonlinear dynamical system) and circumvent the need for linearization in building a quantum algorithm. 

(ii) Another natural possibility is to map the classical nonlinear PDE into an equivalent Schr\"odinger equation. This would mean that the nonlinearity is now encoded into an effective, equivalent Hamiltonian $H_{eq}$, whose dynamics is governed \cite{jin2023quantum} by
\begin{align}
    &i\hbar\frac{\partial\vert\Psi\rangle}{\partial t} =  H_{eq} \vert\Psi\rangle, \label{eq:Schrodinger equation}\\
    \implies &\vert\Psi(t)\rangle = e^{-iH_{eq}t}\vert\Psi(0)\rangle = U_{eq}\vert\Psi(0)\rangle,
\end{align} or by a nonlinear hydrodynamic Schr\"odinger equation \cite{meng2023quantum},  whose solution is essentially the action of a unitary operator $U_{eq}$ that can then be simulated directly on a quantum computer using a Hamiltonian simulation algorithm.

 In the next section, we outline a few quantum algorithms proposed for solving problems that emerge from the above approaches.
 
\subsection{QCFD ALGORITHMS}
\label{subsec:qcfd algorithms}

\subsubsection{Solving Navier-Stokes equations} \label{subsec: solving NS}
Quantum algorithms that consider the NS equations as the starting point have relied on either linearization or on different ansatzes to initialize their input problem setup. Typically, the latter approach leads to an energy/cost function minimization problem, which is solved by a Variational Quantum Algorithm (VQA), in which the ansatz parameters undergo iterative optimizations. The former approach, on the other hand, leads to a problem of linear system of equations, which is solved either by a Quantum Linear Systems Algorithm (QLSA) or by a Variational Quantum Linear Solver (VQLS) algorithm (again solved as a minimization problem). Of the several algorithms proposed in the past, the above two candidates have emerged as promising. We will now highlight certain salient details of these approaches.

\begin{figure}[htpb!]
    \centering
    \includegraphics[trim={0.1cm 0cm 0.1cm 0.1cm},clip=true,scale=0.44]{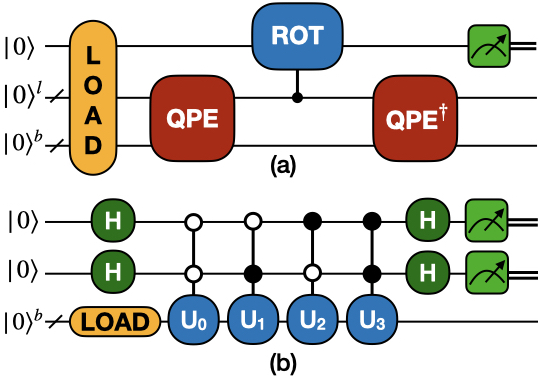}
   \caption{\justifying (a) shows the typical circuit of a HHL algorithm. (b) shows the typical circuit construction a Linear Combination of Unitaries algorithm.}
    \label{fig:qlsa}
\end{figure}

(a) \textit{Quantum Linear Systems Algorithm} (QLSA) --  Matrix operations, especially matrix-vector products and matrix inversions, form a essential part of most CFD schemes for PDEs. Consider the one-dimensional, linear, advection-diffusion equation given by
\begin{equation}
    \frac{\partial u}{\partial t} + {\tilde{U}} \frac{\partial u}{\partial x} = D \frac{\partial^{2} u}{\partial x^{2}},
\end{equation}
where $u$ is the velocity field, ${\tilde{U}}$ is the constant advection velocity and $D$ is the diffusion constant. A central-finite difference discretization of the space and time coordinates, with an explicit and implicit time stepping formulations, yields the following set of equations respectively.
\begin{align}
    &\frac{u_{i}^{j+1}-u_{i}^{j}}{\Delta t} + {\tilde{U}} \frac{u^{j}_{i+1} - u^{j}_{i-1} }{2\Delta x} = D\frac{u^{j}_{i+1}- 2u^{j}_{i} + u^{j}_{i-1} }{(\Delta x)^{2}}, \\
     &\frac{u_{i}^{j+1}-u_{i}^{j}}{\Delta t} + {\tilde{U}} \frac{u^{j+1}_{i+1} - u^{j+1}_{i-1} }{2\Delta x} = D\frac{u^{j+1}_{i+1}- 2u^{j+1}_{i} + u^{j+1}_{i-1} }{(\Delta x)^{2}} 
     .
\end{align}
These equations can be written in the form
\begin{align}
    \textbf{u}^{j+1} &= \tilde{M} \textbf{u}^{j}, \label{eq:explicit}\\
    \textbf{u}^{j+1} &= M^{-1} \textbf{u}^{j},\label{eq:implicit}
\end{align}
respectively. The central task of QLSA algorithms is to compute the solution of these linear system of equations. The first of its kind is the HHL algorithm \cite{harrow2009quantum} that computes the solution of the form, $\textbf{x}=M^{-1}\textbf{b}$. While the corresponding classical algorithm achieves this task, with a complexity of $\mathcal{O}(N\kappa s\log(1/\epsilon))$, the HHL algorithm exhibits an exponential advantage in $N$, with a complexity of $\mathcal{O}(\log(N)\kappa^{2} s^{2}/\epsilon)$. Here, $N$ is the size of the input matrix, $s$ and $\kappa$ are the matrix sparsity and condition numbers, and $\epsilon$ is the error in the solution. The structure of the HHL quantum circuit is shown in figure \ref{fig:qlsa}(a). The first loading step is the quantum state preparation algorithm that encodes the input $\textbf{b}$ in the first register with $\log(N)$ qubits. Subsequently a Quantum Phase Estimation (QPE) step expresses the matrix as $e^{iMt}$, which is then expanded in the eigenbasis of $M$ to compute the binary approximations of the eigenvalues $\Lambda_{j}$. These eigenvalues are then stored in the second register as an $l-$bit, truncated binary number. The conditional rotation step then operates on the third (ancillary) register to compute the inverses $\Lambda^{-1}_{j}$, which together with the first register produces $b_{j}/\Lambda_{j}$. In the eigenbasis, this is the target solution $\textbf{x}$. The QPE$^{\dag}$ resets the second register to $\vert0\rangle^{l}$ and the ancillary register is measured. If the measurement output is $\vert0\rangle$, it implies that the remaining state encodes our desired result. 

While this seems straightforward in principle, its practical implementation is quite involved. To achieve accurate solutions with computational advantage calls for one to transcend many difficulties \cite{aaronson2015read}. They include the sparsity of the matrix, the efficient state preparation algorithm, efficient output measurement, an appropriate $t$ (for $e^{iMt}$) that needs tro be given such that the eigenvalues are scaled appropriately for accurate storage with an $l-$bit approximation, and so on. They are almost never satisfied in their entirety, and is above and beyond the design of the gate level circuit. In the context of flow simulations, ref.~\cite{bharadwaj2023hybrid} proposed solutions to some of these challenges and devised an end-to-end quantum algorithm to achieve accurate and convergent solutions. This work also highlights the importance of implementing the algorithms at a gate level in order to construct a functional quantum solver.

In any case, HHL-type algorithms are typically characterized by deep circuits, not amenable for near-term quantum devices but more suitable for future fault-tolerant devices. This motivated the subsequent QLSA algorithms that brought about crucial improvements \cite{childs2017quantum,subacsi2019quantum,gaitan2020finding,childs2021high,liu2021efficient,an2023quantum,bharadwaj2024compact,berry2024quantum}. Some of these more recent works \cite{an2023quantum,bharadwaj2024compact} are based on a new strategy that amalgamates efficient Hamiltonian simulation algorithms \cite{berry2014exponential,berry2015simulating} and Linear Combination of Unitaries (LCU) method \cite{childs2017quantum}. The former is a set of algorithms that provides an efficient quantum circuit for operators of the form $e^{i\tilde{W}t}$, essentially forming solutions of eq.~(\ref{eq:Schrodinger equation}) for a Hamiltonian $\tilde{W}$. The latter, whose typical circuit is shown in figure \ref{fig:qlsa}(b), is used to approximate an arbitrary matrix operator as a linear combination of unitary operators $U_{i}$, given by $M=\sum\limits_{i=0}^{m}\alpha_{i}U_{i}$, where $\alpha_{i}$ are the corresponding coefficients. For $m=3$, say, the first (ancillary) register that has $\log(m)$ qubits is prepared in a superposition state proportional to $\sum\limits_{i=0}^{3}\vert i\rangle$, via the Hadamard gates. The second register with $\log(N)$ qubits encodes the input state. Then the unitaries $U_{i}$ are applied on the second register using the first register as control qubits. The last set of Hadamard gates then sum the operators to approximate the action of $M$. The solution here is prepared probabilistically, which implies that one would need to measure the solution subspace multiple times to reconstruct the solution. When each of these unitaries are of the form $U_{j}=e^{i\tilde{W}_{j}t}$, for some Hamiltonian $\tilde{W}$, we can again employ the improved Hamiltonian simulation algorithms to build an efficient linear systems algorithm. This forms the underlying tenet of the current state-of-the-art algorithms. These improvements have now culminated into what are known as Linear Combination of Hamiltonian Simulation (LCHS) algorithms \cite{an2023linear}. 

More recently, ref.~\cite{bharadwaj2024compact} proposed a compact quantum algorithm that offers an efficient way to perform matrix operations iteratively for time marching simulations. This algorithm offers a near-optimal complexity (logarithmic or poly-logarithmic) in most system parameters including the grid size $N$, the number of time steps $\tau$, the condition number of the matrix operator $\kappa$ and the accuracy of the solution (except for the sparsity $s$ of the matrix operator). This is complemented by an optimal (logarithmic) qubit complexity, thus making the algorithm one of the front runners in this approach. Apart from having short depth circuits and an end-to-end approach, the algorithm design also accounts for errors from real quantum devices, making it amenable to near-term devices. We highlight here that this ability to perform iterative matrix operations efficiently has a wide range of applications beyond fluid dynamics itself. 

Now we turn to the original nonlinear problem. ref.~\cite{liu2021efficient} employs the algorithm proposed in \cite{childs2017quantum} and appends to it a preceding Carleman linearization, which leads to an algorithm with an overall complexity proportional to $\mathcal{O}(T^{2}{q}\textrm{poly}(\log(TN/\epsilon)/\epsilon)$. Here, $T$ is the time up to which the PDEs are simulated {and $q$ is the measure of decay in the solution with time}. Besides the computational advantage of this algorithm, more importantly, the authors quantify the degree of nonlinearity that is tractable by such a linearization algorithm, by defining a number $R$, analogous to the Reynolds number, $Re$. They show that the algorithm exhibits error convergence (exponentially decaying error) while maintaining the aforementioned complexity, only when $R<1$. This indicates that such an algorithm can only handle weakly nonlinear problems, corresponding to low $Re$ flows. Furthermore, even other methods of linearization, such as Koopman or Homotopy perturbation methods, suggest similar results. Although this might be the case, in the absence of numerical evidence of simulations with a large enough untruncated Carleman terms and without an explicit connection between the two quantities ($R$ and $Re$), it is not possible to concretely comment on the capability of such algorithms to simulate fluid flows. This is an ongoing investigation. These investigations suggest that better bounds might be possible, although they remain to be computed. This observation might raise one's optimism, but we caution that highly nonlinear PDEs of practical interest still demand novel strategies. 

\begin{figure}[htpb!]
    \centering
    \includegraphics[trim={0.1cm 0cm 0.1cm 0.1cm},clip=true,scale=0.44]{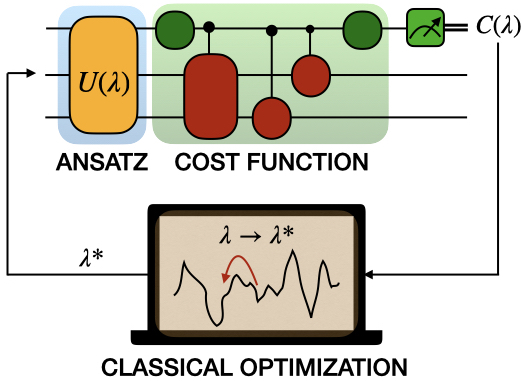}
   \caption{\justifying Shows the schematic workflow of a Variational Quantum Algorithm. The initial input is encoded using the parameterized ansatz $U(\lambda)$. The cost function $C(\lambda)$ which is to be minimized, is then evaluated quantumly and is then output to the classical device. Subsequently, this set of parameters are updated via classical optimization. The new set of parameters $\lambda^{*}$, are then used to compute the cost function again on the quantum device and this process is continued until convergence.}
    \label{fig:vqa workflow}
\end{figure}

\textit{(b) {Variational Quantum Algorithm} (VQA)} -- Another promising avenue is to convert the problem into one of optimization, forming the basis of the Variational Quantum Algorithms~\cite{cerezo2021variational}. The typical workflow of such an algorithm is shown in figure \ref{fig:vqa workflow}. 
The parameterized ansatz $U(\lambda)$ described previously is used to input the data and encode the nonlinearities. The goal is to define and minimize a suitable cost function $C(\lambda)$, which is a function of the input. As shown in \cite{ingelmann2023two}, the residue in the velocity solution of successive times steps \big($C(\lambda)=||~\vert\textbf{u}^{j+1}(\lambda)\rangle-\tilde{M}\vert\textbf{u}^{j}(\lambda)\rangle ||^{2}_{2}$ or = $||~\vert\textbf{u}^{j+1}(\lambda)\rangle-M^{-1}\vert\textbf{u}^{j}(\lambda)\rangle ||^{2}_{2}$\big) produces a converged solution in the limit. The quantum computer is used to accelerate the cost function evaluation, while the parameter updates are done by a classical optimizer algorithm;  the process is repeated until convergence. The quantum circuit for the cost function typically involves computing various terms that emerge from overlaps between guessed and target solutions, for which Hadamard and SWAP tests circuits form indispensable tools. For the classical optimizer, there exists a wide variety of algorithms that include gradient based methods, Nelder-Mead, BFGS, geometric and constrained optimizations, chosen appropriately to the problem being solved~\cite{ingelmann2023two}.

There have been several algorithms proposed so far \cite{lubasch2020variational,leong2022variational,bravo2023variational,ingelmann2023two,wright2024noisy,pool2024nonlinear} to solve both linear and nonlinear flow problems. This approach has several merits that make it an attractive option to pursue:
\begin{itemize}
    \item The algorithm is typically characterized by shallow circuit depths making it amenable to near-term devices.
    \item Encoding the nonlinearity in the PDE is now transformed into a cost function evaluation, which therefore makes handling nonlinearity relatively better compared to linearization techniques, whose closure errors require more qubits to encode large dimensional vector spaces.
    \item The measurement in the quantum algorithm yields only a single real valued output (the cost function), which has a better chance of preserving quantum advantage.
\end{itemize}
While these features seem appealing, there are also certain drawbacks:
\begin{itemize}
    \item Choosing the {ansantz} and right cost function is not straightforward. For certain problems, the size of the parameter space required can grow linearly with the problem size, resulting in deep circuits.
    \item The interplay between parametrization and grid resolution is unclear and generally leads to either under- or over-parameterization {, besides the heuristics that the ansatzes entail. Increasing the accuracy or expressiveness of the ansatz would cause the cost of parameterized state preparation to tend to $N$, thus diminishing any available quantum advantage.}
    \item Based on the specifics of the numerical set up, the classical optimizer may critically slow down and run into convergence issues, despite having sufficient resolution and number of qubits. One again encounters the {barren plateau problem}\cite{holmes2022connecting,larocca2024review}, where the cost function saturates and does not decay identically to 0.
    \item The workaround for these problems are generally heuristic and extremely sensitive to the choice of the problem, the ansatz parametrization and the classical optimization algorithm. For these reasons, the variational approach does not come with complexity guarantees over classical counterparts (unlike QLSA algorithms, for instance). {A remark here is probably useful. Typical algorithms start with a random initial guess (``cold start") to the optimization cycle, which might allow one to have an ansatz whose parameter space is smaller than the problem size $N$. However, one is likely to consider a ``warm start" with a well-informed initial guess to accelerate the optimization process. Such a specific input could increase the required parametrization of the ansatz. Instead, the initial guess could embody certain well-established features of the steady state solution---for instance the skewness of the probability distribution of the velocity field, symmetries in the mean flow characteristics---which could overall aid variational quantum solvers to tackle nonlinear problems.}
    \item{ Although the method appears to be resource efficient, it does not offer a guaranteed quantum advantage.}
\end{itemize}
\begin{figure*}[htpb!]
    \centering
    \includegraphics[trim={0.6cm 6.2cm 0.1cm 0.1cm},clip=true,scale=0.95]{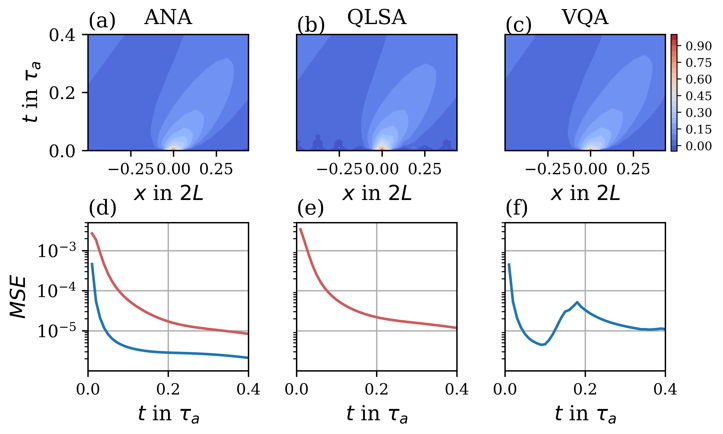}
   \caption{\justifying Contour plots ({reproduced from  J. Ingelmann et al. (2024)}) depict the time evolution of the velocity field from a one-dimensional linear advection-diffusion equation, with $x$ (abscissa) being the spatial dimension and (ordinate) $t$ being the simulation time in normalized units. The diffusion constant $D=1m^{2}/s$, system size $L=1m$ and constant advection velocity of ${\tilde{U}}=10m/s$. (a) Shows the analytical solution where as (b) and (c) shows the solutions computed by QLSA and VQA algorithms for a system discretized with $N=16$ grid points and a time step size of $\Delta t =0.0001s$}
    \label{fig:vqa_qlsa_ana}
\end{figure*}
An example of a simulation result can be seen in figure \ref{fig:vqa_qlsa_ana}, which shows that both QLSA and VQA algorithms can accurately capture the analytical result. The QLSA, on one hand, offers rigorous complexity guarantees on computational advantage, without heuristics and barren plateau problems. Recent efforts such as ref.~\cite{bharadwaj2024compact} address state preparation and measurement issues, as well as how to lower the circuit depths to sizes comparable to variational approaches. However, the step of encoding nonlinearities is more involved with truncation errors and circuit sizes needed for full-blown simulation sizes need more resources compared to variational methods.

The balancing act involving this list (not comprehensive) makes it unclear at this stage whether either (both or neither) of them would emerge as the winner for future quantum simulations of flow problems. Such an algorithm will likely combine the strengths of both methods.

\subsubsection{Solving Liouville equation} \label{subsec: solving liouville}
An alternative starting point would be to solve the linear Liouville equation (\ref{eq:liouville}) or the Fokker-Planck equation (\ref{eq:fokker planck}) that circumvents need for linearization. Solving this equation reduces to resorting to QLSA or VQA methods. This implies that the solution brings with it the strengths and weaknesses (discussed above) of the algorithm one chooses, but there is no need for linearization and multiple ansatz copies. However, some studies \cite{lin2022koopman,succi2024ensemble} seem to suggest that while this route may seem impeded by the following bottlenecks:
\begin{itemize}
    \item Computing the final velocity field solution assumes the availability of an accurate closure scheme for the probability distribution in the phase space. This is generally not given and difficult to compute. This might be regarded as the manifestation of truncation errors in linearization methods.
    \item When one tries to improve the numerical resolution by refining the grid, the resulting solutions are infected with spurious oscillations (Gibbs effect). This issue cannot be resolved without nonlinear feedback.
    \item This method requires one to perform an averaging of these stochastic PDEs by making a large ensemble of simulations, which increases the required computational resources many fold.
\end{itemize}
Without addressing these setbacks, the potential of this approach remains unclear. 
\subsubsection{Solving the Lattice Boltzmann equation} \label{subsec: solving LBM} 
\begin{figure}[htpb!]
    \centering
    \includegraphics[trim={0cm 0cm 0.1cm 0.1cm},clip=true,scale=0.35]{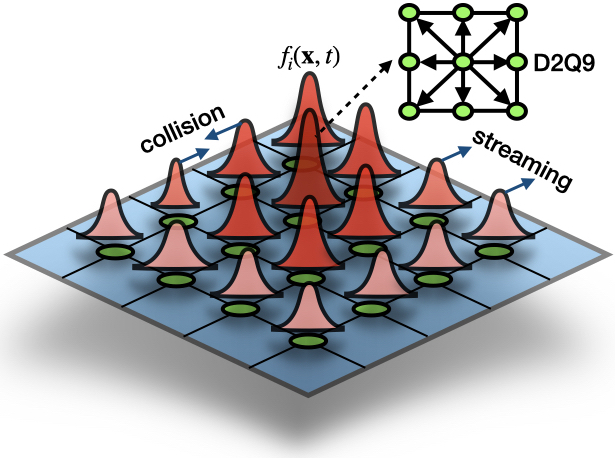}
   \caption{\justifying 2 dimensional Lattice Boltzmann discretized lattice with a D2Q9 designated velocity directions for the density distributions $f_{i}(\textbf{x},t)$.}
    \label{fig:lbm}
\end{figure}

In contrast to the continuum scale approach discussed so far, a meso-scale formulation is given by the Lattice Boltzmann equations, which are discretized versions of the Boltzmann equation solved on a lattice, as shown in figure \ref{fig:lbm}. A density distribution $f_{i}(\textbf{x},t)$ is assigned to every lattice site that moves around the lattice with a pre-designated velocity ($\textbf{v}_{i}$) in every direction. A $d-$dim system has a total of $3^{d}$ hopping directions and designated velocities (written as D1Q3, D2Q9 and D3Q27, respectively). The evolution of these distributions is governed by the Lattice Boltzmann equation
\begin{equation}
 f_{i}(\textbf{x}+\textbf{v}_{i}\Delta t,t+\Delta t) - f_{i}(\textbf{x},t) = -\frac{\Delta t}{\tau}((f_{i}(\textbf{x},t)-f^{eq}_{i}(\textbf{x},t),
\end{equation}
where $\Delta t$ is the discrete time step, $\tau$ is the relaxation time and $f^{eq}_{i}$ is the equilibrium density distribution. This evolution can be split into two important processes shown in figure \ref{fig:lbm}: collision and streaming. They are given respectively by
\begin{align}
    f_{i}(\textbf{x},t+\Delta t)&=f_{i}(\textbf{x},t)-\frac{\Delta t}{\tau}((f_{i}(\textbf{x},t)-f^{eq}_{i}(\textbf{x},t),\label{eq:collision}\\
   f_{i}(\textbf{x},t+\Delta t)&=\tilde{f}(\textbf{x}+\textbf{v}_{i}\Delta t,t+\Delta t)\label{eq:streaming}.
\end{align}
The goal of a quantum algorithm in this context is to efficiently simulate these two processes as a quantum circuit. Typically, the density information corresponding to the lattice site position and momentum are encoded as binary values using the quantum basis states. Nevertheless, an amplitude encoding is possible as well. While streaming is a linear operation, collision is nonlinear. There have been several efforts \cite{todorova2020quantum,ljubomir2022quantum,li2023potential,fonio2023quantum,kocherla2023fully,schalkers2024efficient,sanavio2024lattice,itani2024quantum,frankel2024quantum} so far that propose quantum algorithms to perform these tasks using different approximations. These include collisionless models, exact/inexact streaming, unitary/non-unitary collision and streaming, bit-shift and bit-arithmetic methods, and so on. Each of these methods have their respective strengths and weaknesses, which we shall not consider in detail here. It suffices to highlight here that, if we were to solve the fully nonlinear problem efficiently, we require a linearization procedure such as the Carleman technique that leads to a linear system of equations. The resulting problem can then be solved by simulating the streaming and collision steps by using either a native quantum circuit as shown in ref.~\cite{itani2024quantum} or by using QLSA as in ref.~\cite{li2023potential}. Comparing all these different approaches that have been proposed so far, it appears that the method that eventually uses QLSA will possess a net quantum advantage. This, however, does not come with guarantees on the degree of nonlinearity it can handle. Nevertheless, if we are allowed to consider quantum advantage as a reasonable measure, the QLSA based method might be a regarded as strong contender.

Comparing this method with the continuum formulation, we note that both approaches deal with nonlinearity (except for VQA) via a linearization technique, which limits one to weakly nonlinear problems. Further, the data encoding as amplitudes in the continuum approach makes them more qubit efficient compared to the lattice approach and boundary conditions are more complicated to handle. In any case, even if these shortcomings were to be eliminated, the resulting algorithm would still have the same complexity as that of a QLSA algorithm, while possessing the disadvantage of a meso-scale approach over its continuum counterparts.

\subsubsection{Solving Schrödinger equation} \label{subsec: solving schrodinger}More recently, another approach that has emerged in context of solving flow problems is by converting them into an equivalent Schr\"odinger equation that evolves under an equivalent Hamiltonian. This can be done in two ways:

 \textit{(a) Madelung Transform} -- In ref.\cite{meng2023quantum} the authors show that by using what is known as a Madelung transformation, the Navier-Stokes equations can be approximated as a nonlinear Hydrodynamic Schr\"odinger equation. These efforts have relied on tools such as Hamiltonian simulations, Quantum Fourier Transforms, Phase estimations, as well as certain custom circuit designs. The nonlinearity in this case is not handled explicitly by the quantum device, but instead relegated to a prediction-correction method. We remark that this approach seems more natural for a quantum device. However, Madelung transforms are typically for a inviscid potential flows and therefore efforts to find tighter and exact mappings of a fully viscous Navier-Stokes equations is required. Also, a computational advantage of the overall algorithm still remains theoretically elusive at present.  
    
\textit{(b) Schr\"odingerization} -- More recently works such as refs.~\cite{jin2023quantum,lu2023quantum,hu2024quantum} have worked towards developing a more exact mapping between classical PDEs and an equivalent Schr\"odinger equation by using a technique known as the \textit{warped phase transform}. This approach is well suited for continuous-variable quantum computing where, instead of a discrete representation of qubits, one uses what are known as qumodes. These qumodes encode an infinite-dimensional space of continuous values.  Therefore a $d-$dimensional discrete problem is mapped to a $\hat{d}+1$-dimensional analog representation. The extra dimension comes from the warp-transform parameter $\xi$, that is given by $\hat{u}(x,t,\xi)=e^{-\xi}u(x,t)$. This mapping yields an exact representation of the input classical PDE as a Schr\"odinger equation with an equivalent Hamiltonian $H_{eq}$. The algorithm performs Hamiltonian simulation by implementing operators of the form $e^{-iH_{eq}t}$.  

Given this structure, this approach enjoys certain upsides. Since a continuous variable approach is used, discretization of PDEs is no longer required, and the Hamiltonian and algorithm specific operators are more natural to a quantum devices. Both deterministic and stochastic linear PDEs can be mapped by this approach. However, in comparison to the previous methods, in the absence of a linearization method, the present approach relies on either level-set transformations (similar to Cole-Hopf methods for Burgers equation) or exact transformations. However, such transformations are rare and remain elusive for general nonlinear PDEs such as the Navier-Stokes equation. For the continuous variable version, one needs to possess the hardware that is capable of handling such operations. Optical experiments with coherent states and Boson sampling methods might appear as good candidates in this regard. Finally, comparing one-to-one, the complexity of this approach over the traditional qubit and gate-based quantum algorithms does not appear to be favourable.

\subsubsection{Auxiliary quantum acceleration and simulators} 
\label{subsec: auziliary accelerators}
Constant progress is being made in different avenues mentioned thus far with a quest for a quantum algorithm that, either fully quantumly or as a hybrid method, solves flow PDEs efficiently. It is also worth looking at quantum computers as auxiliary accelerator modules instead of a full blown PDE solver. We mean that the flow problem is primarily solved on a classical device to obtain the velocity field, but a segment of the classical algorithm or a post-processing step of a classical solver is done quantumly to accelerate the overall process. To better convey the sense, let us keep aside the practical challenges for a moment and consider the Quantum Fourier Transform. It is a valuable tool that could accelerate classical, pseudo-spectral simulations of turbulence by replacing the expensive FFTW calculations, which switches the representation of the solution between real and spectral spaces. Along similar lines, one could consider solving the Linear Stability Analysis problem using QLSA tools mentioned above. 

In the context of fluid flow simulations, there have been a few isolated contributions such as in ref.~\cite{pfeffer2023reduced}, where the authors solve a convection flow problem classically but use a quantum algorithm for implementing a machine learning method known as reservoir computing. They attempt to predict flow solutions and dominant dynamical modes in the flow. Assuming data can be encoded efficiently, one could also think of using quantum computer as a post processing unit, where a Quantum Post Processing algorithm such as the one proposed in ref.~\cite{bharadwaj2023hybrid} or Quantum Image Processing tools~\cite{wang2022review} can be used to analyse classical data efficiently. On a slightly different note, classical algorithms can also be redesigned using lessons from quantum computing to build ``quantum-inspired" algorithms~\cite{gourianov2022quantum} that solve certain problems classically, but more efficiently than its own classical predecessors. In these proposals, the quantum computer is only used as an accelerator. Given that these accelerators tend to require only a nominal set of resources, they have great potential of being experimentally realized on near-term quantum devices.

\subsubsection{QFlowS} As important as it is to theorize new quantum algorithms, it is critical to actually implement the algorithm by addressing every detail to verify the algorithm's correctness and performance. They reveal the true capacity of the algorithm and the actual amount of resources needed in practice. Therefore, in the absence of quantum devices that can simulate large problem sizes, classical simulators of quantum algorithms become extremely important. To this end, we have built a high performance Quantum Flow Simulator called QFlowS 
~\cite{bharadwaj2024qflows,bharadwaj2023hybrid,bharadwaj2024compact,ingelmann2023two}, which is a versatile tool that can be used to design, build and test new quantum algorithms by either constructing new quantum circuits from the gate level, or using a library of in-built quantum algorithms. Designed with a matrix-free approach, QFlowS can simulate circuits with up to 30 qubits and a few million gates. This is complemented with an in-built library of classical, computational fluid dynamics tools that can be readily used in conjunction with the quantum algorithms. Thus, QFlowS forms a useful test-bed for future QCFD simulations and algorithm development. As a follow up step, one can use other commercially available tools such as Qiskit (IBM) to translate the algorithms for implementation on real quantum devices.

\section{Veracity of quantum advantage}
\label{sec: veracity of advantage}
Quantum advantage refers to the scenario in which a quantum computer can \textit{accurately} solve a problem \textit{more efficiently} than a classical computer. In accuracy, a given quantum algorithm performing a task with an accuracy $\epsilon$, should in other respects be comparable to a classical algorithm admitting the same accuracy. By efficiency, one does not merely mean comparing the wall-clock times; in the context of fluid dynamics simulations, one measure would be given by how the computational resources scale with increasing problem size ($N$). A crucial remark is that quantum advantage is truly achieved when the overall quantum algorithm--- considering all aspects of quantum state preparation, quantum state tomography and measurements as well as effects of noise and decoherence on actual devices---achieves accuracy and efficiency. Several algorithms proposed so far claim quantum advantage, after relying on intermediate black-boxes and oracles that don't necessarily have efficient circuit implementations. Furthermore, the many caveats of quantum algorithms tend to get brushed aside, obscuring the true challenges that lie within actual implementations of a quantum algorithm, let alone achieving quantum advantage. It is therefore all the more important that the community turn towards what we call end-to-end algorithms, which we briefly outline below. A better sense of this can be had from ref.~ \cite{bharadwaj2024compact}.
 \begin{figure}[htpb!]
    \centering
    \includegraphics[trim={0.1cm 0.1cm 0.1cm 0.1cm},clip=true,scale=0.32]{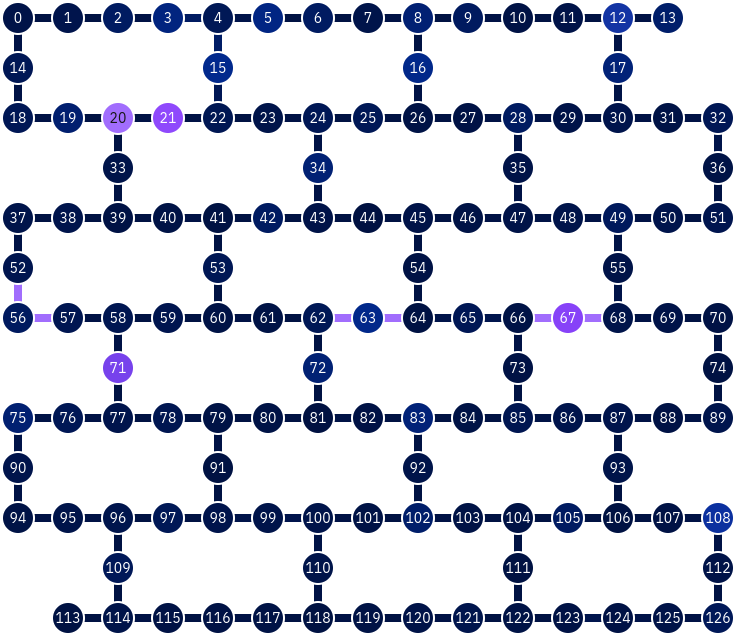}
   \caption{\justifying The qubit topology of a 127 qubit IBM Sherbrooke Eagle-r3 processor. The colors on every qubit indicates the magnitude of readout assignment errors, with the darker shades corresponding to lower errors. The interconnects and their colors represent gate error rates.}
    \label{fig:sherbrooke}
\end{figure}
\vspace{-0.8cm}
\subsection{END-TO-END ALGORITHMS}
\label{subsec: end-to-end}
 In the context of a QLSA approach, an attempt was made in ref.~\cite{bharadwaj2024compact}, similar to a VQA counterpart in ref.~\cite{wright2024noisy}, where ideal simulations, noisy simulations as well as experiments on a real quantum device were performed.

\textit{(a) Quantum State Preparation} -- There exists quantum state preparation algorithms that can efficiently load quantum states with features such as sparse data~\cite{mozafari2022efficient}, functional forms such polynomials and convex integrable functions ~\cite{gonzalez2024efficient,grover2002creating}, specific forms such as normal distributions, skewed distributions and delta functions~\cite{vasquez2022enhancing,rattew2021efficient,iaconis2024quantum}, etc. Efficient algorithms also exist if we admit probabilistic state preparations or if we are willing to trade in more qubits for short depth circuits. For variational algorithms this role is assumed by the choice of the ansatz.
    
\textit{(b) Measurements and Quantum Post Processing} -- An efficient state preparation is alone insufficient to preserve quantum advantage, if the final measurement process is expensive ($\mathcal{O}(N)$) and erroneous. Variational algorithms are in some sense protected from this bottleneck since they generally always measure only the cost function which is typically a single, real-valued output. For all other algorithms, measuring the final state of the simulation is critical. Firstly, for iterative time marching solvers, measuring the entire velocity field solution at every time step of the simulation compromises quantum advantage. By using methods proposed in ref.~\cite{bharadwaj2024compact} one can circumvent this bottleneck by either computing solutions for all time steps in one step or compute the solution at given final time. Even after this, measuring the final state completely is expensive. An alternative would be measure a specific function of the velocity field such as the mean viscous dissipation, mean kinetic energy, higher order gradients of the field, etc. The post processing algorithm in ref.~\cite{bharadwaj2023hybrid} can be used to compute linear and nonlinear functions of the field to output a single real-valued output that characterizes the velocity field. For extracting linear functions and properties of the quantum state such as entanglement entropy, shadow tomography algorithms proposed in ref.~\cite{huang2020predicting} appear to be quite promising.

Beyond this, one should also recall that the final solution is always present as a probabilistic state. By this we mean that to actually reconstruct even a single qubit state, one would need to repeat the circuit execution multiple times, followed by measurements, in order to construct a statistics of the probability distribution of the quantum state amplitudes. These repeated runs are called shots ($N_{s}$), or also the query complexity. The overall time complexity of an algorithm is the gate complexity multiplied by $N_{s}$. Methods such as Richardson extrapolation proposed in ref.~\cite{bharadwaj2024compact} can be used to lower the required shot counts. In practice, we require this overall time complexity to fare better than a classical algorithm.  
     
\textit{(c) Error correction \& circuit optimization} -- The final test is to run the designed algorithm on a real quantum device and successfully maintain the prescribed accuracy and quantum advantage. However, as already mentioned in Section 3.4, decoherence and noise on current devices form a major bottleneck in simulating large quantum circuits with admissible accuracy. 

(i) \textit{Decoherence }-- To combat errors due to decoherence, one needs to construct quantum circuits that are shallow enough for the execution times to be shorter than $T_{1}$ and $T_{2}$ (in case of superconducting qubits). For the IBM Sherbrooke processor shown in figure \ref{fig:sherbrooke}, the median values for these (as on June 24, 2024) were $275.72\mu s$ and $160.63\mu s$ respectively. While variational algorithms generally offer short-depth circuits, even QLSA methods can be carefully optimized to yield shallow circuits as shown in ref.~\cite{bharadwaj2024compact} to meet these requirements. The decoherence times on each qubit keeps deteriorating with time and algorithmic usage. This can happen even across the several shots of execution of a single quantum circuit simulation. These processors are therefore re-calibrated periodically to restore them to their original states. This implies that simulations need to be averaged over different times of a day and over different days to account for these variations.\\
(ii) \textit{Gate errors and logical qubits} -- To ameliorate these errors, one requires special error mitigation and error correction algorithms. A simple algorithm would be to use Richardson extrapolation, where one runs simulations at two distinct and finite noise levels and are used to extrapolate to the zero noise limit \cite{temme2017error}. Error correction circuits are generally appended between circuits to keep errors under check. However, if the gate errors are high, the error correction circuits add more error to the circuit.  Refs.~\cite{aliferis2005quantum,gottesman2022opportunities} hypothesized that, when the gate error rate is below a certain threshold ($\sim 10^{-5}$), the error correction algorithms can kick in, enabling one to simulate arbitrarily deep quantum circuits with little error. To achieve these error rates and reliable qubits that protect the stored information, we need what are known as \textit{logical qubits}. Every logical qubit is made of more than one physical qubit, all of which store a copy of the same information. Noise might affect some of them, if not all. Now on this mixed bag of qubits, by performing what are known as \textit{syndrome measurements}, one can recover the original information. On a noisy gate on a logical qubit, the information would be protected thereby, lowering error rates. More recently, independent efforts by IBM and Quantinuum teams have brought down error rates in superconducting qubits to less than $10^{-7}$. Ref.~\cite{bharadwaj2024compact} suggests that these are actually the error rates required to solve a flow problem by the QLSA algorithm proposed there. Furthermore, \cite{bharadwaj2024compact} also hints towards an interesting proposal of using noise to advantage. In any case, every logical qubit with such error rates are made up of about 10 to 25 physical qubits. This is critical in assessing the qubit resources required for QCFD algorithms. Furthermore, it is also important to consider the qubit topology of the quantum processor while designing algorithms. {On commercially available quantum platforms such as IBM's Qiskit, one can use built-in tools such as coupling maps to allow us to select specific qubits with the least instantaneous errors and a specific topology in physical space, conforming to the entangling gate operations in the quantum circuit.}\\

A successful end-to-end quantum simulation of a flow problem needs to go beyond each of these challenges in order to maintain quantum advantage. 
       
\section{Discussion and outlook}
\label{sec: discussion}
After a brief introduction to quantum computing and the merits claimed for it, we presented a discussion of its challenges, highlighting the need to think ``quantumly" to gain from the advances. We then highlighted several important QCFD algorithms that have been proposed so far, their core strategies, strengths, weaknesses as well as some remedies proposed for handling nonlinearity. We discussed the veracity of quantum advantage, its several caveats and the need for end-to-end algorithms to actually realize a full fledged quantum simulation. We remark here that the present paper is by no means a detailed study of all algorithms mentioned but presents merely an impression. 
However we still make some broad remarks here. 

For demonstrating the utility of quantum computing in applied fields such as fluid dynamics, it is critical that the algorithms are actually implemented as quantum circuits starting at the gate level, to ascertain the correctness of the algorithm. The gap between theoretical proposals and actual implementations is wider than it appears. Bridging this gap is an important task, as is designing new algorithms. 
Besides such challenges as quantum state preparation and measurements, the one of nonlinearity is the most severe. Even amongst attempts made so far, it is unclear whether there is a clear victor. Quantum computers operate on vector spaces that grow exponentially which, in some sense, is a result of quantum entanglement that exists in a nonlinear many-body interacting system. Quantum computers are \textit{nonlinear} in this sense, although the units of their operations are linear. Given this, it is possible that this issue of nonlinearity may not be fundamentally insurmountable. Even if it remains as one, the possibility of quantum computers to accelerate simulations of nonlinear PDEs remains a possibility in an auxiliary sense. In any case, it is important for the community to focus on either improving existing approaches or discovering novel strategies to efficiently and accurately encode nonlinearity. This includes algorithms that can compute nonlinear functions of the resulting velocity solutions. 

Designing quantum algorithms for fault-tolerant quantum devices is critical. At the same time, in order to make progress in the near-term, one needs to begin designing and simulating quantum circuits by posing constraints on circuit depths, gate and shot counts, noise and decoherence, using specifications of current quantum devices. This inverse approach is critical in discovering efficient near-term strategies, which depend on the quantum computer realization that one aims to work with \cite{tennie2024quantum}. For instance, such an exercise could lead to the exploration of noise to our advantage \cite{bharadwaj2024compact}.
This should always be complemented by an end-to-end approach, where every step of overall classical-quantum algorithm is accounted for and implemented in such a way that quantum advantage is not compromised along the way. 

\begin{figure}[htpb!]
    \centering
    \includegraphics[trim={0.1cm 0cm 0.1cm 0.1cm},clip=true,scale=0.28]{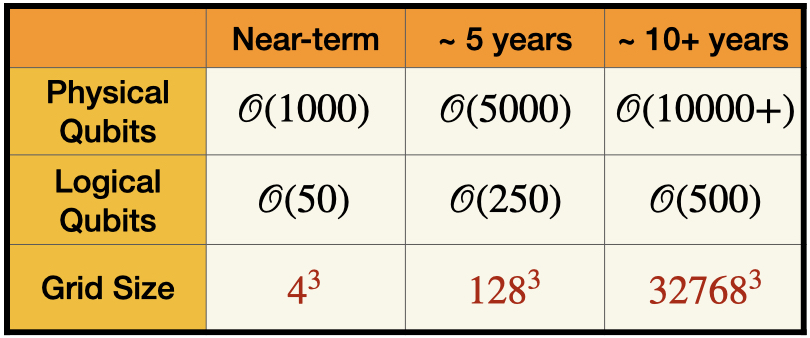}
    \caption{\justifying Roadmap of QCFD simulation sizes for $\tau=10^{6}$ time steps. Note that these numbers are only rough estimates, keeping the timeline of superconducting qubits research (IBM and Quantinuum) as reference. We consider that the number of physical qubits required for every logical qubit to $\in [10,25]$. }
    \label{fig:roadmap}
\end{figure}

On the whole, we believe that simulations of fluid flows using quantum computers has a promising future. In figure \ref{fig:roadmap}, we take IBM's roadmap of superconducting qubits as reference and estimate the simulation sizes (grid size) that become possible in the future. As a disclaimer, these numbers are specific to mostly superconducting qubits, and can differ for other realizations. It appears that the simulation of turbulence on grid sizes of a million would need at least, about 500 logical qubits. Even though a simulation of such size may be achieved, the highest $Re$ of such a simulation is difficult to predict. If the quantum advantage of simulating high $Re$ flows can be established even theoretically, it would be a milestone result.

Finally, we highlight that, for sustained progress in transforming quantum computing into an actual utilitarian tool, physicists, mathematicians and engineers need to work together in bridging the gap between theory and application, while also training the workforce specialised in this area.


\vspace{-1.1cm}
{\section*{Acknowledgements}

The authors wish to acknowledge insightful discussions with various colleagues, particularly Jörg Schumacher, Balu Nadiga, Dhawal Buaria, Philipp Pfeffer, Julia Ingelmann and Wael Itani.}

\section*{End note: A few personal remarks on Professor Roddam Narasimha (RN)}

We are pleased to dedicate this article to RN's memory. He passed away before quantum computing became a subject of interest---certainly well before its application for solving fluid dynamical problems was considered, but we believe that he would have taken serious interest in the subject had he lived longer. Those of us who knew him and his scientific breadth will appreciate the truth of this remark. For instance, he championed parallel computing and was instrumental in developing India's first, Flosolver Mark 1, a parallel computing hardware \cite{narasimha2021rocket}. It would have intrigued him that the unitary and linear structure of quantum computing could probably be used with advantage for solving nonlinear fields such as those governed by the Navier-Stokes equations. While, as discussed in the main body of the article, much needs to happen before this goal becomes a reality, there is a glimmer of hope that it would be possible in the not-too-distant future. Is it a competition for the classical digital computing? By no means, at this stage, but who can tell what the future holds!

KRS recalls with pleasure the four or five years he spent with RN as a graduate student; of all the things he learnt then, the most enduring lesson was to be open to learning new things. SSB spent about 1.5 years doing research with RN. Both of us remember him as a scholar of great breadth, inquisitive and enthusiastic about new developments---and as a gentleman in all personal interactions. The legacy that RN has left behind is immense, only a part of which is visible in the few articles collected here.

We thank Professor Jaywant Arakeri who, as editor of Sadhana, put his weight behind this collection of articles honoring RN. 




\end{document}